\documentclass{gCMB2e}

\usepackage{subfigure}
\usepackage{amssymb}
\usepackage{amsmath}
\usepackage{graphicx}

\begin{document}





\title{Fluid dynamics of heart valves during atrial fibrillation: a lumped parameter-based approach}

\author{S. Scarsoglio$^{\rm a}$$^{\ast}$\thanks{$^\ast$Corresponding author. Email: stefania.scarsoglio@polito.it
\vspace{6pt}}, C. Camporeale$^{\rm b}$, A. Guala$^{\rm b}$ and L. Ridolfi$^{\rm b}$\\\vspace{6pt} $^{a}${\em{Department of Mechanical and Aerospace Engineering, Politecnico di Torino, Torino, Italy}};\\
$^{b}${\em{Department of Environment, Land and Infrastructure Engineering, Politecnico di Torino, Torino, Italy}}}

\maketitle

\begin{abstract}
Atrial fibrillation (AF) consequences on the heart valve dynamics are usually studied along with a valvular disfunction or disease, since in medical monitoring the two pathologies are often concomitant. Aim of the present work is to study, through a stochastic lumped-parameter approach, the basic fluid dynamics variations of heart valves, when only paroxysmal AF is present with respect to the normal sinus rhythm (NSR) in absence of any valvular pathology. Among the most common parameters interpreting the valvular function, the most useful turns out to be the regurgitant volume. During AF both atrial valves do not seem to worsen their performance, while the ventricular efficiency is remarkably reduced.


\end{abstract}

\section{Introduction}

Atrial fibrillation (AF), with accelerated and irregular beating, is the most widespread cardiac arrhythmia, yielding important symptoms (such as palpitations, chest pain, shortness of breath, reduced exercise ability) and decreasing the cardiac and energetic performances \citep{Fuster}. Besides the overall impact on the cardiovascular system, it is of interest from a lifestyle and clinical point of view understanding how AF modifies the heart valve fluid dynamics.

Valvular dynamics during AF is important, as AF is potentially a new risk marker for mitral insufficiency \citep{Enriquez-Sarano2010}. Moreover, patients with aortic regurgitation are at higher risk if subject to AF \citep{Dujardin}. Both left and right atrial remodeling due to long-term AF aggravate mitral and tricuspid regurgitation \citep{Yamasaki,grigioni}. Two main reasons explain why clinical results analyzing the role of AF alone on the valvular dynamics are difficult to be found. First, accurate echocardiographic measurements and different levels of approximation are further needed to quantify the valvular function. For example, \cite{Zhou} evaluate the grade of regurgitation severity only through the valve orifice area. Second, since AF rarely comes up without side pathologies, its net contribution to the hemodynamic changes is hardly identifiable. In fact, clinically unrecognized moderate mitral regurgitation is strictly correlated, either as an etiologic factor or a consequence, to lone AF \citep{Sharma}. Moreover, patients with mitral regurgitation and AF seem to benefit from the restoration of the normal sinus rhythm, as AF worsens the valvular function when mitral insufficiency is already present \citep{Gertz}. But what happens to the transvalvular flow dynamics due to the sole presence of AF without the additional role of valvular diseases remains poorly investigated and represents the focus of the present work.

Regurgitant fraction, regurgitant and forward volumes are common parameters typically adopted to classify the degree of valvular insufficiency diseases. Their estimate usually involve Doppler echocardiographic measurements \citep{Touche}, cardiac magnetic resonance, and cardiac computed tomography \citep{Thavendiranathan}. In most of the cases, however, further geometry-based approximations (such as the Proximal Isovelocity Surface Area, PISA, for the mitral valve, \cite{Enriquez-Sarano1995}) are needed because the regurgitant volume can only be computed as difference between mitral inflow and aortic outflow stroke volume in the absence of other diseases or intracardiac shunts \citep{Thavendiranathan}. Owing to the difficulty of calculating the regurgitant volume (and therefore the regurgitant fraction), peak mitral inflow velocity has been proposed as an alternative measure for grading the severity of the valvular disease. Even if a positive correlation between maximum peak values and regurgitant fraction emerges, limitations of this approach can arise during physiologic and fibrillated beating \citep{Ozdemir,Thomas}.

The lumped-parametrization modelling here proposed - and already used to evaluate the global impact of AF on the cardiovascular system \citep{scarsoglio2014} and study the role of heart rate during AF \citep{anselmino2015} - allows us to obtain the complete flow rate temporal series for the four valves. In this way, performing a statistical analysis over thousands of heart beats, we are able to quantify with no approximations all the most relevant valvular parameters: minimum and maximum flow rate peak values (which, through the orifice area, are proportional to the velocity peaks), as well as regurgitant and forward volumes, and therefore the regurgitant fraction. By means of these parameters and the valve opening time intervals, we can describe the main valvular changes deriving from a condition of paroxysmal AF alone. 
Surprisingly, AF seems to improve the atrial valve efficiency in terms of regurgitant fraction, while the ventricular valve dynamics undergo a substantial deterioration.

\section{Materials and Methods}

\subsection{Mathematical Modelling and Beating Features}

The lumped parameter model here used to evaluate the valve dynamics extends the Windkessel approach to the complete cardiovascular system, including the systemic and venous circuits together with an active representation of the four cardiac chambers, and consists of a network of compliances, resistances, and inductances. The cardiovascular system is described in terms of volumes, $V$ [ml], pressures, $P$ [mmHg], flow rates, $Q$ [ml/s], and valve opening angles, $\theta$ [$^\circ$]. Each of the four chambers is governed by an equation for the mass conservation (accounting for the volume variation, d\emph{V}/d\emph{t}), a constitutive equation (relating pressure, $P$, and volume, $V$, through a time-varying elastance, $E$), an orifice model equation for the pressure-flow rate relation, and an equation for the valve motion mechanisms. The systemic and pulmonary circuits are divided into five sections, four for the arterial circulation and a unique compartment for the venous return. Each section is ruled by an equation of motion (accounting for the flow variation, d\emph{Q}/d\emph{t}), an equation for the conservation of mass (expressed in terms of pressure variations, d\emph{P}/d\emph{t}), and a linear state equation between pressure, $P$, and volume, $V$. The present model has been compared with more than thirty clinical studies giving an overall good agreement in terms of the hemodynamic response during AF \citep{scarsoglio2014}.

\noindent The valve dynamics has recently received new insights from the mathematical modelling (see, among many others, \cite{baccani,Aboelkassem,Weinberg}). Here several valve motion mechanisms are considered, such as the pressure difference across the valve, the frictional effects from neighboring tissue resistance, the dynamic motion effect of the blood acting on the valve leaflet, the action of the vortex downstream of the valve, while the shear stress on the leaflet is very small and is neglected \citep{Korakianitis}. The external forces acting on the leaflets of each valve result in a second-order differential equation for each opening angle, $\theta$. All the physical mechanisms introduced rely on previous 2D or 3D CFD analyses, which focused on local flow features (please refer to \cite{Korakianitis}, and related references therein). Thus, although the three-dimensional flow dynamics are not directly inserted into the model, all the most relevant 3D effects, such as the vortex downstream of the valve, are reliably modeled and accounted for through the lumped parametrization. The heart dynamics description provides to model valvular regurgitation and the dicrotic notch mechanism.

\noindent The resulting ordinary differential system is numerically solved through an adaptative multistep numerical scheme for stiff problems implemented in
Matlab by the \texttt{ode15s} function. The algorithm is based on the numerical differentiation formulas of variable order (from 1st to 5th order). As the system shows some stiffness features, in particular when rapid variations occur in correspondence of the valve opening and closing, the adopted solver proves to be very efficient and suitable especially in handling the valve dynamics (for details, refer to \cite{scarsoglio2014}). To compare normal sinus rhythm (NSR) and AF outcomes, 5000 cardiac cycles are simulated in both configurations to guarantee the statistical stationarity of the results.

Normal RR heart beats are extracted from a correlated pink Gaussian distribution (mean: $\mu=0.8$ s, standard deviation: $\sigma=0.06$ s), representing a physiologic heart rate of 75 bpm. AF beatings are instead extracted from an Exponentially Gaussian Modified distribution (mean $\mu=0.67$ s, standard deviation: $\sigma=0.17$ s), which is the most common AF distribution ($60-65\%$ of the cases, more details can be found in \cite{scarsoglio2014}). As AF inhibits atrial contraction, both atria are maintained passive in this configuration.

\subsection{Definition of parameters}

We here introduce parameters used to evaluate the flow dynamics of each valve. The forward volume, $FV$ [ml/beat], is defined as the volume of blood per beat flowing forward through the valve:

\begin{equation}
FV = \int_{RR} Q^+(t) dt,
\end{equation}

\noindent where with the symbol $Q^+$ we indicate the positive flow rate outgoing from the valve. The regurgitant volume, $RV$ [ml/beat], is the volume of blood per beat which regurgitates backward through the valve:

\begin{equation}
RV = \int_{RR} Q^-(t) dt,
\end{equation}

\noindent where the symbol $Q^-$ represents the negative flow rate going backward through the valve ($RV<0$ by definition).  The combination of the forward volume and the regurgitant volume is the net flow through the valve, known as stroke volume, $SV=FV+RV$ [ml/beat]. The regurgitant fraction, $RF$, is the percentage of flow regurgitating back through the valve with respect to the flow in the forward direction, $RF = |RV|/FV \cdot 100 \%$. The regurgitant volume and fraction are standard measures grading the valve regurgitation severity. When the $RV$ and $RF$ values are lower than 30 ([ml/beat] and $\%$, respectively), regurgitation is considered mild. When instead $RV>60$ ml/beat and $RF>50\%$, there is a severe regurgitation. Moderate regurgitation has values in between those of mild and severe \citep{Zoghbi,Cawley}. $T_f$ [s] is the temporal interval over which the valve is open with forward flow, while $T_b$ [s] indicates the time lapse characterized by backward flow. These temporal ranges are more meaningful if presented and discussed as percentages of the RR interval: $\tau_f = T_f/RR \cdot 100\%$ and $\tau_b = T_b/RR \cdot 100\%$.

In Fig. 1a, the forward volumes of mitral and aortic valves are shown in light colors, while dark colors reproduce the regurgitant volumes for a single normal sinus rhythm beat (RR=0.8 s). Maxima and minima values are also indicated (please note that during AF the A wave, representing the atrial kick, is absent), as well as the temporal ranges, $T_f$ and $T_b$. An analogous representation is reported for the right heart valves in Fig. 1b.

\section{Results}

The main results for NSR and AF are summarized in Table 1 in terms of mean and standard deviation values computed over 5000 cycles. The net stroke volume per beat, $SV$, is constant throughout the four valves, with a reduction when passing from NSR (63.8 ml/beat) to AF (53.6 ml/beat). The $SV$ decrease ($-16\%$), usually observed during AF (please see related references in \cite{scarsoglio2014}), is mainly driven by the RR shortening (NSR: $\mu=0.8$ s, AF: $\mu=0.67$ s), as the mean flow rate is about 80 ml/s for all valves in both NSR and AF configurations. Regurgitant fraction and volume values fall within the physiologic range \citep{Kalmanson,Sechtem,Roes} for both NSR and AF cases. Standard deviation values of all the parameters reported in Table 1 are higher during AF with respect to NSR, as a consequence of the higher variability of the fibrillated RR beating (NSR: $\sigma=0.06$ s, AF: $\sigma=0.17$ s). To better highlight how AF can differently alter the valve dynamics and flow rate repartition, results are discussed focusing on one valve at a time (blue: NSR, red: AF). Maximum peak correlations with $RF$ in normal and fibrillated conditions are reported in Appendix A for the four valves, while in Appendix B details on the valve opening angle dynamics during  NSR and AF are discussed.

\subsection{Mitral valve}

Minimum peak values are deeper during AF (NSR: -289.6 ml/s, AF: -357.4 ml/s), while regurgitant volume is almost halved (NSR: -10.0 ml/beat, AF: -5.8 ml/beat). This can be explained by the fact that, during AF, the temporal interval for backward flow, $\tau_b$, is about half of the NSR case (NSR: 8.2 \%, AF: 4.4 \%). Since fibrillated atria are passive, during atrial systole there is no atrial kick and the valve at the end of this phase smoothly approaches the closure (d\emph{Q}/d\emph{t} is much smaller than during the atrial kick, with values close to zero), as observable in Fig. 2a.

Maximum peak values remain averagely constant comparing NSR (923.8 ml/beat) and AF (920.6 ml/beat). Forward volume undergoes a substantial decrease (NSR: 73.9 ml/beat, AF: 59.4 ml/beat) because during AF A-wave is missing and the mean HR increases. Since $RV$ decreases more than $FV$ does when passing from NSR to AF, this leads to a decay of the regurgitant fraction (NSR: 13.6 \%, AF: 9.9 \%).

Low minimum peak values correspond to small $RF$ as evidenced in Fig. 2b. Concerning the relation between maximum peak values, $Q_{max}$, and $FV$, there is a positive correlation in NSR, which becomes sparse in AF for higher $FV$ values (see Fig. 2c). Minimum peak values and $RV$ present a sparse relation (see Fig. 2d) during NSR, which turns into a weak positive correlation in AF.

\subsection{Aortic valve}

The concomitant deepening of minimum peak values (NSR: -242.3 ml/s, AF: -307.1 ml/s) and lengthening of the temporal interval in backward flow, $\tau_b$, (NSR: 4.8 \%, AF: 5.9 \%), leads to an increase of the regurgitant volume during AF (NSR: -5.3 ml/beat, AF: -6.5 ml/beat). Maximum peak values remain constant (NSR: 1268.5 ml/s, AF: 1265.4 ml/s), while the temporal range for forward flow, $\tau_f$, slightly increases (NSR: 22.2 \%, AF: 23.6 \%). However, this does not prevent the reduction of the forward volume (NSR: 69.1 ml/beat, AF: 60.1 ml/beat). In fact, during AF, once the maximum peak value is reached the flow rate usually experiences a more rapid deceleration (please refer to Fig. 3a) which decreases the forward volume (with respect to NSR having the same valve opening time). The forward volume decrease of $13\%$  and the regurgitant volume increase of $24\%$ lead to a $44\%$ increase for the $RF$ (NSR: 7.6 \%, AF: 10.9 \%).

Deeper minimum peak values are directly related to higher $RF$ (see Fig. 3b). Maximum peak values are barely related to the amount of $FV$ (see Fig. 3c), while there is a clear linear proportion between minimum peak values and $RV$ (Fig. 3d). This relation becomes piecewise linear during AF: the lower branch emerges as a result of the accelerated beating (RR$<$0.5 s, see red symbols of Fig. 3d).

\subsection{Tricuspid valve}

Minimum peak values modestly vary (NSR: -264.3 ml/s, AF: -244.3 ml/s) and the temporal interval in backward flow, $\tau_b$, reduces significantly (NSR: 9.5 \%, AF: 6.1 \%). Similarly to the mitral case, the $\tau_b$ reduction during AF is mainly due to the missing valve opening contribute activated by the atrial kick (which is absent also in the right atrium, see Fig. 4a). Therefore, the $RV$ is almost halved (NSR: -10.2 ml/beat, AF: -5.6 ml/beat), while the absence of atrial kick and the increased mean HR also imply an average reduction of the $FV$ during AF (NSR: 74.0 ml/beat, AF: 59.2 ml/beat). As a result, the $RF$ is substantially decreased in AF (NSR: 13.8 \%, AF: 9.7 \%).

Minimum peak values are deeper for larger $RF$ values (see Fig. 4b). However, minimum peak values are scarcely correlated with $RV$ values during NSR, while a positive correlation emerges in AF (see Fig. 4d). An unexpected negative correlation relates maximum peak values and $FV$ (see Fig. 4c). When longer beats occur, maximum peak values decrease but, after the peak, the forward flow presents a region where flow rate is almost constant until the valve is open, thereby enhancing the forward volume (please consider time series in Fig. 4a). When, on the contrary, rapid beating emerges maximum peaks are higher and the plateau regions are almost absent, resulting in a lower forward volume. This behaviour is already noticeable in NSR but is much more evident during AF (see Fig. 4c, red symbols), where the RR variability is higher.

\subsection{Pulmonary valve}

During AF minimum peak values are deeper (NSR: -117.3 ml/s, AF: -158.5 ml/s) and the time lapse in backward flow, $\tau_b$, increases (NSR: 5.8 \%, AF: 7.4 \%). As a consequence, the regurgitant volume is higher (NSR: -3.1 ml/beat, AF: -4.2 ml/beat). Despite a modest increase of maximum peak values (NSR: 842.9 ml/s,  AF: 856.3 ml/s) and temporal range in forward flow (NSR: 28.3 \%,  AF: 30.1 \%), the forward volume decreases by $14\%$ (NSR: 67.0 ml/beat, AF: 57.8 ml/beat). As in the aortic valve, after the maximum peak flow rate is reached, there is a much more rapid deceleration during AF, which in turn reduces the forward volume (see Fig. 5a). The regurgitant fraction, mainly due to a $RV$ increase of $35\%$, grows by $59\%$ (NSR: 4.7 \%, AF: 7.5 \%).

Higher $RF$ values are accompanied by deeper minimum peak values (see Fig. 5b). Maximum peak values are scarcely correlated to forward volumes (Fig. 5c), instead a remarkable linear proportion holds minimum peaks and $RV$ (see Fig. 5d), as happened for the aortic valve.

\section{Discussion}

For all the valves, the temporal range in forward flow, $\tau_f$, is poorly influenced by irregular and accelerated beating (mitral NSR: 45.4 \%, AF: 46.1 \%; aortic NSR: 22.2 \%, AF: 23.6 \%; tricuspid NSR: 49.9 \%, AF: 50.6 \%; pulmonary NSR: 28.3 \%, AF: 30.1 \%). The time interval over which the valve is open in backward flow, $\tau_b$, is instead more prone to vary during AF (mitral NSR: 8.2 \%, AF: 4.4 \%; aortic: NSR: 4.8 \%, AF: 5.9 \%; tricuspid: NSR: 9.5 \%, AF: 6.1 \%; pulmonary: NSR: 5.8 \%, AF: 7.4 \%), and this variation makes the $RV$ contribution on the resulting $RF$ more relevant than $FV$. In fact, if $\tau_b$ increases (or decreases), $RV$ in absolute terms grows (or decays). An analogous dependance is not valid between $\tau_f$ and $FV$: $\tau_f$ weakly increases from NSR to AF, while $FV$ is always damped.

Aortic and pulmonary valves undergo similar changes one to each other. Although in the physiologic range, ventricular valves face an important flow dynamics worsening due to a substantial increase of the $RF$ during AF (aortic from NSR to AF: + 43 \%; pulmonary from NSR to AF: + 59 \%), which is largely imputable to the increased temporal interval of backward flow, $\tau_b$. On the contrary, both mitral and tricuspid valves diminish their $RF$ values during AF ($RF$: mitral from NSR to AF: -27 \%; tricuspid from NSR to AF: -36 \%), as the absence of atrial kick reduces the subsequent time range in backward flow. In all valves, $FV$ values decrease during AF, either as a consequence of the missing atrial contraction (atrial valves) or due to the more rapid deceleration of maximum peaks (ventricular valves). Therefore, what makes the difference in terms of $RF$ is the regurgitant volume behaviour, which for ventricular valves grows, while for atrial valves decreases.

Apart from the tricuspid valve, AF deepens minimum peak values. Regarding maximum peaks, AF increases the right heart ones, while the left heart peaks remain practically unaltered. Moreover, AF makes maximum ventricular flow rates steeper. Therefore, the global effect is to accentuate the instantaneous flow rate (maximum and minimum) peaks, without having a correspondent and univocal increase of (forward and regurgitant) volumes. As previously mentioned, indeed, $RV$ and $FV$ values are intrinsically linked to the valve opening ranges, too. In the aortic and pulmonary valves, $\tau_b$ increases during AF, thus a clear positive correlation is found between minimum peak values and $RV$. For the mitral and tricuspid valves, $\tau_b$ is reduced and no distinguishable trend relates $Q_{min}$ and $RV$.

\section{Limitations}

The present model is limited to analyze paroxysmal AF events, as long-term structural remodeling effects due to the chronic persistence of AF are not taken into account. Due to the complexity of the mathematical details, the coronary circulation is neglected as well. Short-term regulation feedbacks of the baroreceptor mechanism, which acts to partially contrast the beating irregularity and acceleration, are absent.

\noindent Concerning the heart rate stochastic modelling, we focused on the most common RR distribution, which is unimodal, as representative of the fibrillated beating. Different less prevailing multimodal RR distributions are not considered here.

\section{Conclusions}

The main changes on the valvular fluid dynamics during AF are driven by the temporal interval over which backward flow occurs. For this reason, the parameter which, by itself, can better indicate the variations of the flow rate cardiac performance is the regurgitant volume. In fact, on one hand, maximum peaks are not meaningful to predict the regurgitant fraction in physiologic and fibrillated conditions. On the other hand, minimum peaks positively correlate with $RF$, but a mean deepening (or shortening) of regurgitant peaks does not necessarily imply an averagely higher (or lower) $RF$.

The present study focused on the role of AF alone on the heart valve dynamics, a thing which is rarely accomplished in clinical practise, where side pathologies complicate the medical framework. In particular, the cardiovascular system here analyzed is not affected by any valvular dysfunction or disease. Aortic and pulmonary flow rate performances become less efficient during AF. Surprisingly, AF does not seem to worsen the efficiency of both atrial valves. 

\section*{Appendix A. Maximum peak flow rates and regurgitant fraction}

In Fig. A1 maximum peak flow rates, $Q_{max}$, of the four valves are contrasted with the corresponding $RF$ values. Based on the present results, the relation linking maximum peak values and $RF$ does not seem to be appropriate to predict the $RF$ severity in both NSR and AF. In the mitral case, in particular, higher $RF$ values are associated to lower maximum peaks (and vice versa), especially during AF (see Fig. A1, panel a). For the remaining valves, even if a weak positive correlation is distinguishable, data are quite sparse (see Fig. A1, panels b to d). As already observed, the maximum peak velocity is usually used to grade the level of valvular diseases, while is not very useful for predicting mild/physiologic degrees of $RF$ and is unable to interpret the severity of $RF$ in patients with atrial fibrillation \citep{Ozdemir,Thomas}.

\section*{Appendix B. Opening angle dynamics for atrial and ventricular valves}

\subsection*{Atrial valves}

When the atrial kick is absent (that is, in AF), the atrial valves during the atrial systole do not increase their opening angles, $\theta$, but gradually reach a lower value (see Fig. B1 panels a and c, where the valve angle, $\theta$, is reported for NSR in blue and AF in red for the mitral and tricuspid valves, respectively). Starting from a lower angular value, the closure interval is more rapid, thereby reducing the temporal range, $\tau_b$, over which backward flow is allowed, as well. In fact, we recall that regurgitant backflow occurs during the closing phase. The reason for earlier atrial valves opening during AF is that the absence of atrial kick decreases the ventricular pressures, so that during diastole it is easier for the atrial pressures to exceed the ventricular pressures. In NSR, the valves - which had slowly started to close approaching the atrial systole - are pushed again to the maximum opening value by means of the atrial kick (see Fig. B1 panels a and c, blue curves) lengthening the closing phase, a thing which in turn increases the interval, $\tau_b$, over which the flow is in backward direction.

\subsection*{Ventricular valves}

The absence of atrial kick do not substantially change the ventricular opening angle dynamics. At the end of the ventricular systole the ventricular valves reach a bit higher opening angles in AF. Starting from a higher angular value, the closure interval is less rapid, therefore $\tau_b$ is increased for the ventricular valves (see Fig. B1 panels b and d, where the valve angle, $\theta$, is reported for NSR in blue and AF in red for the aortic and pulmonary valves, respectively). The earlier valve opening during AF is due to a decrease of the (systemic and pulmonary) aortic pressures, so that during the ventricular systole the ventricular pressures are more facilitated to exceed the aortic pressures.

\clearpage

\begin{table}
\begin{center}
\begin{tabular}{|c|c|c|c|}
  \hline
  & & NSR & AF \\
  \hline
  & $Q_{max}$ [ml/s] & 923.8 $\pm$ 8.5 & 920.6 $\pm$ 17.6 \\
  \cline{2-4}
  & $Q_{min}$ [ml/s] & -289.6 $\pm$ 13.9 & -357.4 $\pm$ 77.1 \\
  \cline{2-4}
  & $FV$ [ml/beat] & 73.9 $\pm$ 2.6 & 59.4 $\pm$ 5.9 \\
  \cline{2-4}
  Mitral & $RV$ [ml/beat] & -10.0 $\pm$ 0.1 & -5.8 $\pm$ 1.5 \\
  \cline{2-4}
  Flow & $RF$ [\%] & 13.6\% $\pm$ 0.5 & 9.9\% $\pm$ 2.9 \\
  \cline{2-4}
  & $\tau_{f}$ [\%] & 45.4\% $\pm$ 1.7 & 46.1 \% $\pm$ 6.3 \\
  \cline{2-4}
  & $\tau_{b}$ [\%] & 8.2 \% $\pm$ 0.2 & 4.4 \% $\pm$ 0.8 \\
  \hline
  & $Q_{max}$ [ml/s] & 1268.5 $\pm$ 19.4 & 1265.4 $\pm$ 92.4 \\
  \cline{2-4}
  & $Q_{min}$ [ml/s] & -242.3 $\pm$ 23.1 & -307.1 $\pm$ 72.9 \\
  \cline{2-4}
  & $FV$ [ml/beat] & 69.1 $\pm$ 2.2 & 60.1 $\pm$ 6.2 \\
  \cline{2-4}
  Aortic & $RV$ [ml/beat] & -5.3 $\pm$ 0.5 & -6.5 $\pm$ 1.3 \\
  \cline{2-4}
  Flow & $RF$ [\%] & 7.6\% $\pm$ 0.9 & 10.9\% $\pm$ 2.8 \\
  \cline{2-4}
  & $\tau_{f}$ [\%] & 22.2 \% $\pm$ 0.6 & 23.6 \% $\pm$ 2.7 \\
  \cline{2-4}
  & $\tau_{b}$ [\%] & 4.8 \% $\pm$ 0.3 & 5.9 \% $\pm$ 1.2 \\
  \hline
  & $Q_{max}$ [ml/s] & 830.7 $\pm$ 9.9 & 860.9 $\pm$ 38.3 \\
  \cline{2-4}
  & $Q_{min}$ [ml/s] & -264.3 $\pm$ 8.6 & -244.3 $\pm$ 30.3 \\
  \cline{2-4}
  & $FV$ [ml/beat] & 74.0 $\pm$ 3.2 & 59.2 $\pm$ 9.9 \\
  \cline{2-4}
  Tricuspid & $RV$ [ml/beat] & -10.2 $\pm$ 0.3 & -5.6 $\pm$ 0.9 \\
  \cline{2-4}
  Flow & $RF$ [\%] & 13.8 \% $\pm$ 0.6 & 9.7 \% $\pm$ 2.3 \\
  \cline{2-4}
  & $\tau_{f}$ [\%] & 49.9 \% $\pm$ 1.6 & 50.6 \% $\pm$ 6.1 \\
  \cline{2-4}
  & $\tau_{b}$ [\%] & 9.5 \% $\pm$ 0.3 & 6.1 $\pm$ 1.1 \\
  \hline
  & $Q_{max}$ [ml/s] & 842.9 $\pm$ 14.6 & 856.3 $\pm$ 63.4 \\
  \cline{2-4}
  & $Q_{min}$ [ml/s] & -117.3 $\pm$ 10.9 & -158.5 $\pm$ 45.6 \\
  \cline{2-4}
  & $FV$ [ml/beat] & 67.0 $\pm$ 2.9 & 57.8 $\pm$ 8.8 \\
  \cline{2-4}
  Pulmonary & $RV$ [ml/beat] & -3.1 $\pm$ 0.3 & -4.2 $\pm$ 1.2 \\
  \cline{2-4}
  Flow & $RF$ [\%] & 4.7 \% $\pm$ 0.6 & 7.5 \% $\pm$ 2.4 \\
  \cline{2-4}
  & $\tau_{f}$ [\%] & 28.3 \% $\pm$ 0.8 & 30.1 \% $\pm$ 3.6 \\
  \cline{2-4}
  & $\tau_{b}$ [\%] & 5.8 \% $\pm$ 0.4 & 7.4 \% $\pm$ 1.6 \\
  \hline
\end{tabular}
\end{center}
\caption{Valvular parameters in terms of mean and standard deviation computed over 5000 cardiac cycles. I column: NSR, II column: AF.}
\end{table}

\clearpage

\section*{List of Captions}

\section*{Figures}

\noindent Fig. 1: (a) Aortic ($Q_{ao}$, red) and mitral ($Q_{mi}$, blue) flow rates, and (b) pulmonary ($Q_{po}$, red) and tricuspid ($Q_{ti}$, blue) flow rates for a typical normal beat (RR=0.8 s). Light colors represent forward volumes ($FV$), dark colors indicate regurgitant volumes ($RV$).

\bigskip

\noindent Fig. 2: Mitral flow, $Q_{mi}$, NSR: blue, AF: red. (a) Representative temporal series. (b) Regurgitant fraction, $RF$, as function of $Q_{min}$. (c) Forward volume, $FV$, as function of $Q_{max}$. (d) Regurgitant volume, $RV$, as function of $Q_{min}$.

\bigskip

\noindent Fig. 3: Aortic flow, $Q_{ao}$, NSR: blue, AF: red. (a) Representative temporal series. (b) Regurgitant fraction, $RF$, as function of $Q_{min}$. (c) Forward volume, $FV$, as function of $Q_{max}$. (d) Regurgitant volume, $RV$, as function of $Q_{min}$.

\bigskip

\noindent Fig. 4: Tricuspid flow, $Q_{ti}$, NSR: blue, AF: red. (a) Representative temporal series. (b) Regurgitant fraction, $RF$, as function of $Q_{min}$. (c) Forward volume, $FV$, as function of $Q_{max}$. (d) Regurgitant volume, $RV$, as function of $Q_{min}$.

\bigskip

\noindent Fig. 5: Pulmonary flow, $Q_{po}$, NSR: blue, AF: red. (a) Representative temporal series. (b) Regurgitant fraction, $RF$, as function of $Q_{min}$. (c) Forward volume, $FV$, as function of $Q_{max}$. (d) Regurgitant volume, $RV$, as function of $Q_{min}$.

\bigskip

\noindent Fig. A1: Regurgitant fraction, $RF$, as function of $Q_{max}$ (NSR: blue, AF: red): (a) mitral, (b) aortic, (c) tricuspid, (d) pulmonary.

\bigskip

\noindent Fig. B1: Valve opening angles, $\theta$ [$^\circ$], during a typical NSR beat (RR=0.8 s, blue) and an AF beat (RR=0.67 s, red): (a) mitral, (b) aortic, (c) tricuspid, (d) pulmonary.

\clearpage

\begin{figure}
\includegraphics[width=\columnwidth]{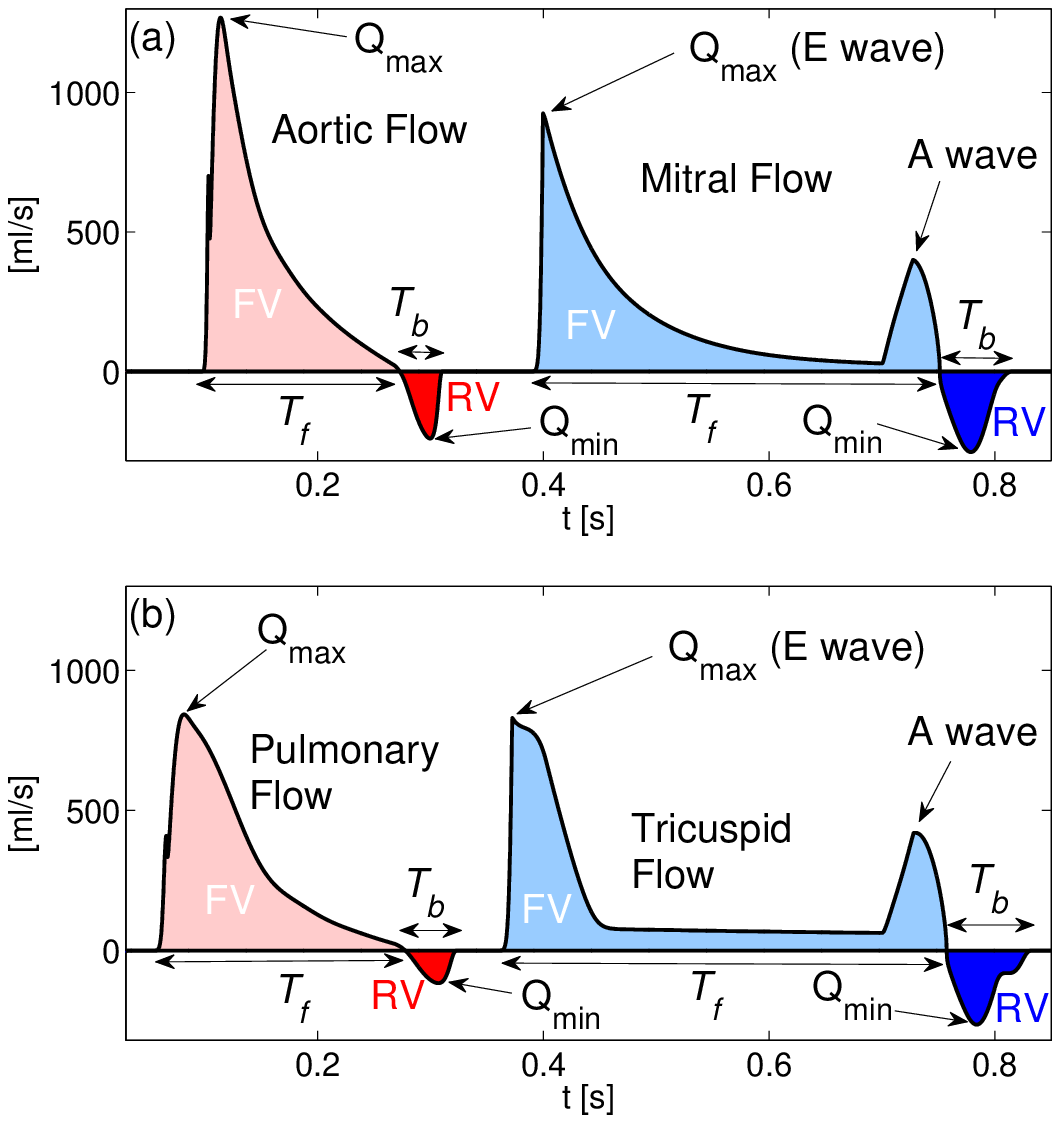}
\label{scheme}
\caption{}
\end{figure}

\clearpage

\begin{figure}
\includegraphics[width=\columnwidth]{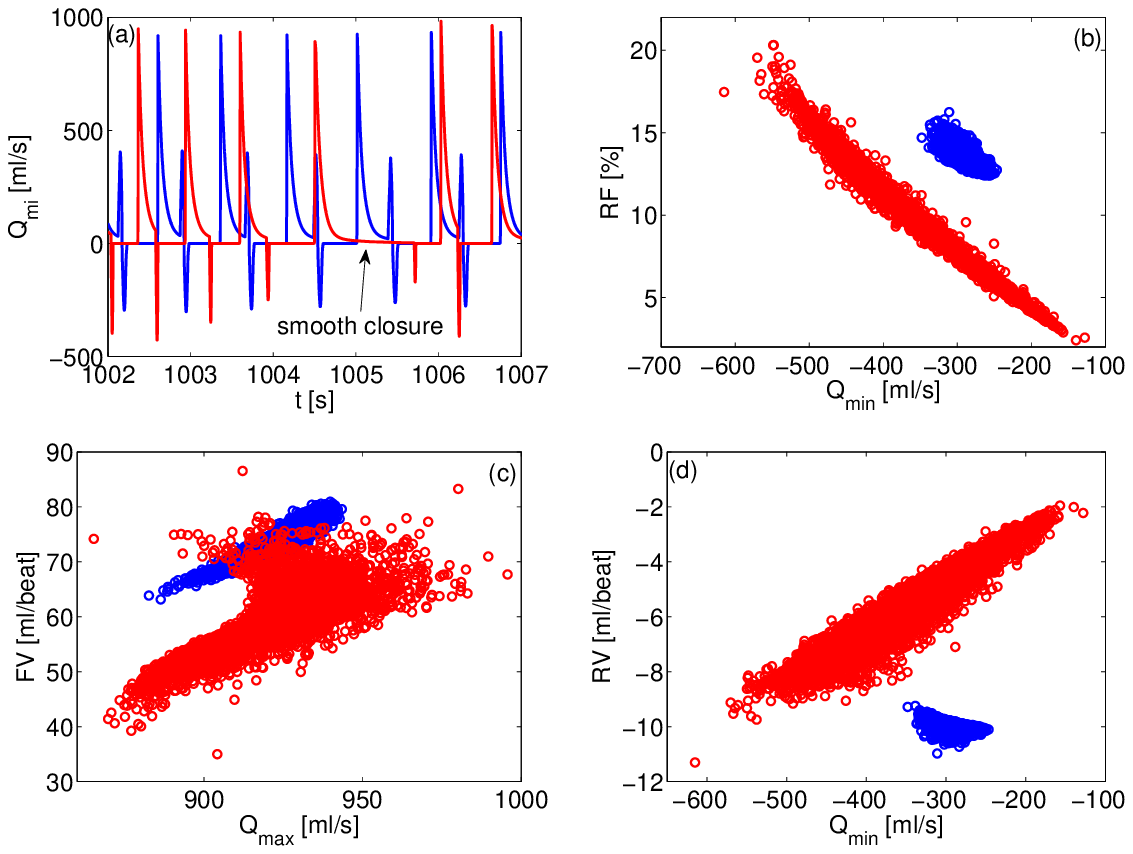}
\label{Qmi}
\caption{}
\end{figure}

\clearpage

\begin{figure}
\includegraphics[width=\columnwidth]{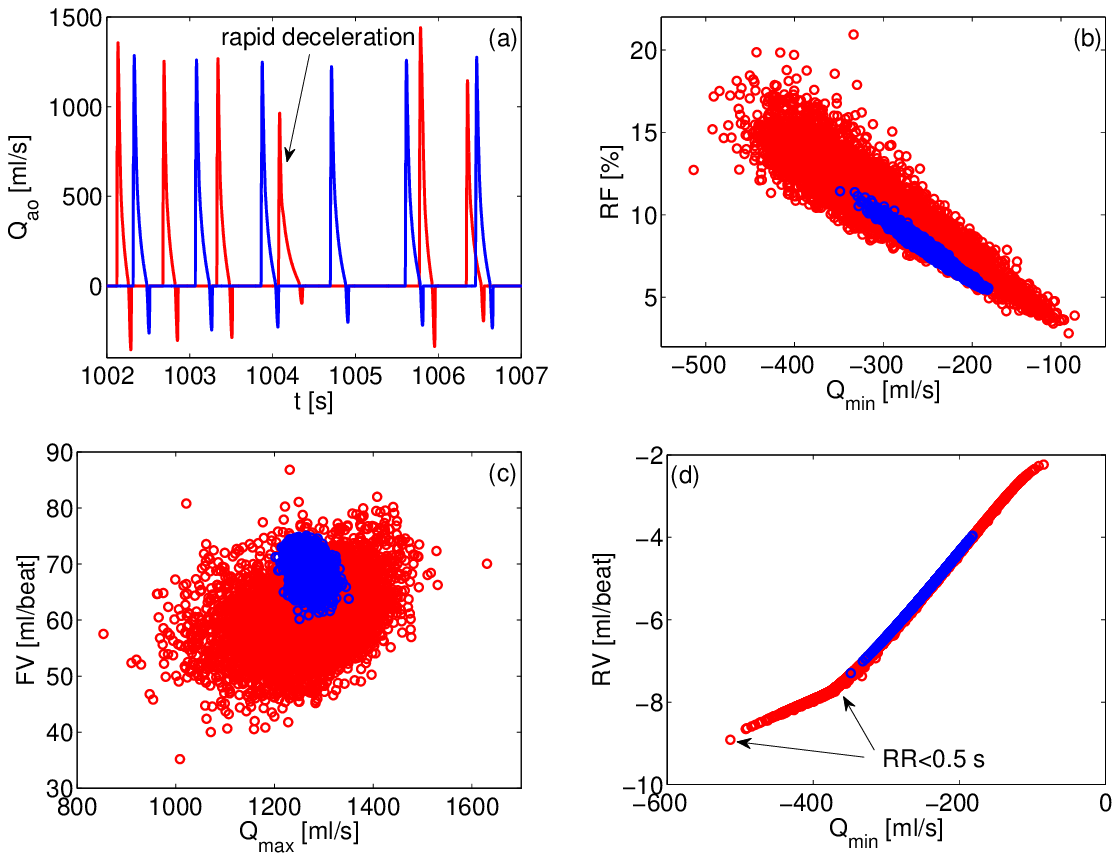}
\label{Qao}
\caption{}
\end{figure}

\clearpage

\begin{figure}
\includegraphics[width=\columnwidth]{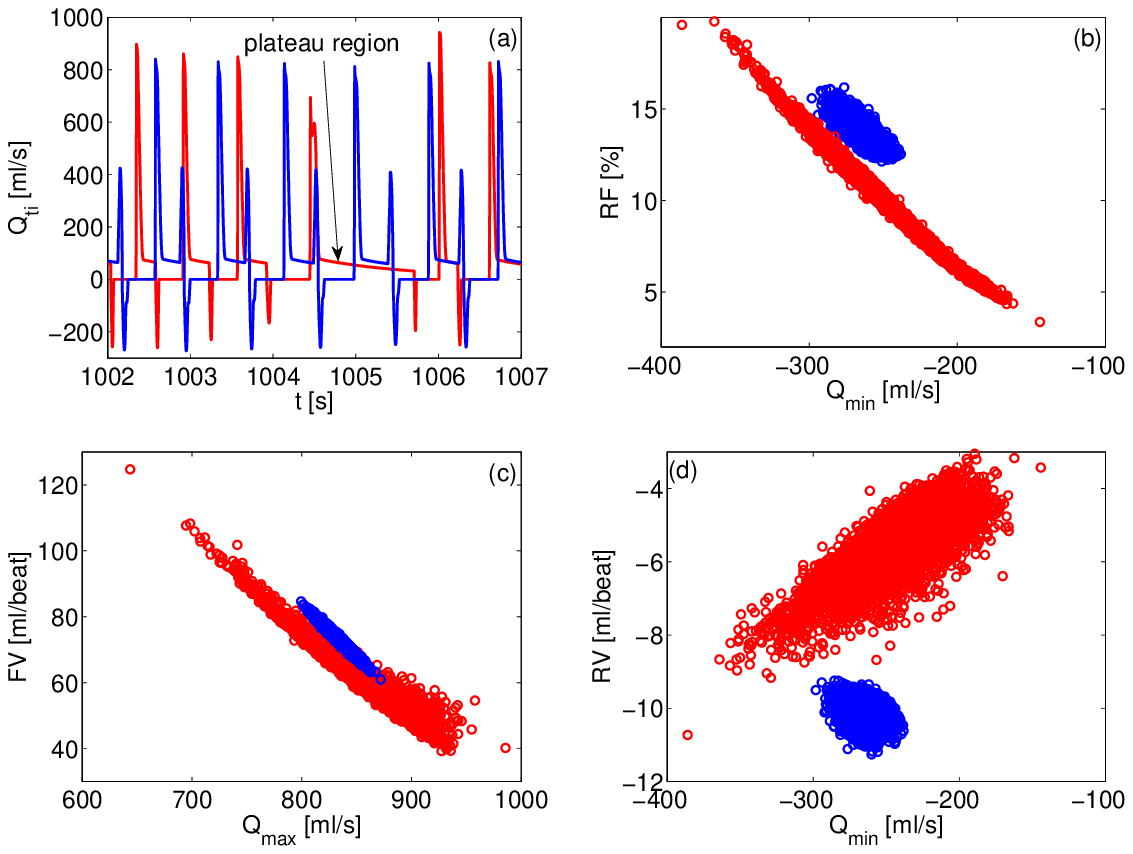}
\label{Qti}
\caption{}
\end{figure}

\clearpage

\begin{figure}
\includegraphics[width=\columnwidth]{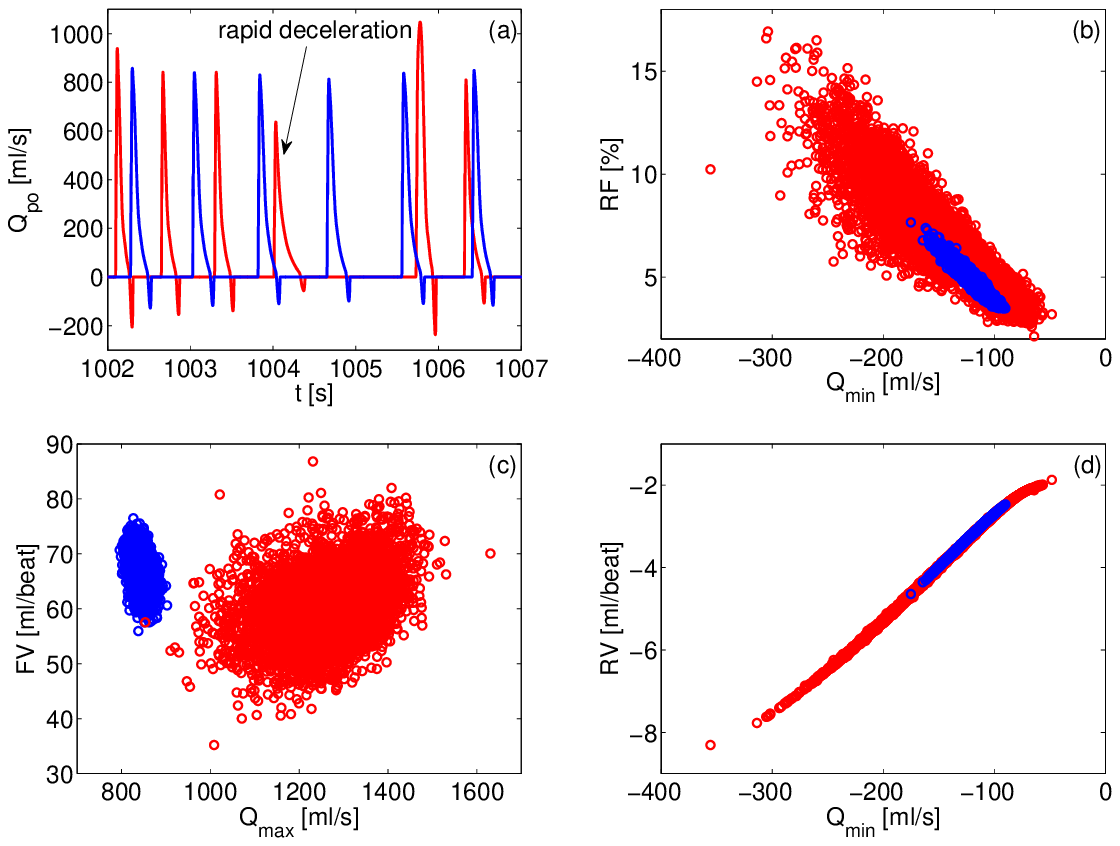}
\label{Qpo}
\caption{}
\end{figure}

\clearpage

\begin{figure}
\includegraphics[width=\columnwidth]{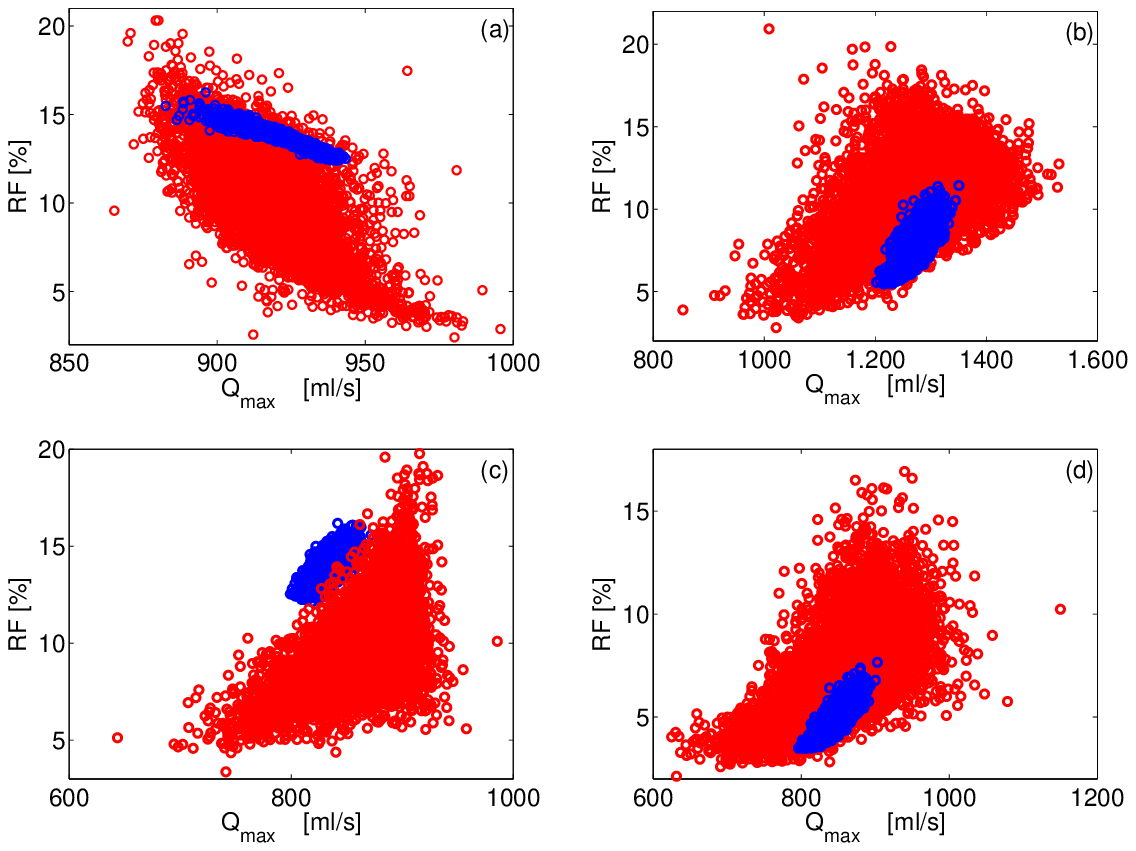}
\renewcommand{\caption}{\footnotesize Figure A1.}
\caption{}
\label{Appendix_A}
\end{figure}

\clearpage

\begin{figure}
\includegraphics[width=\columnwidth]{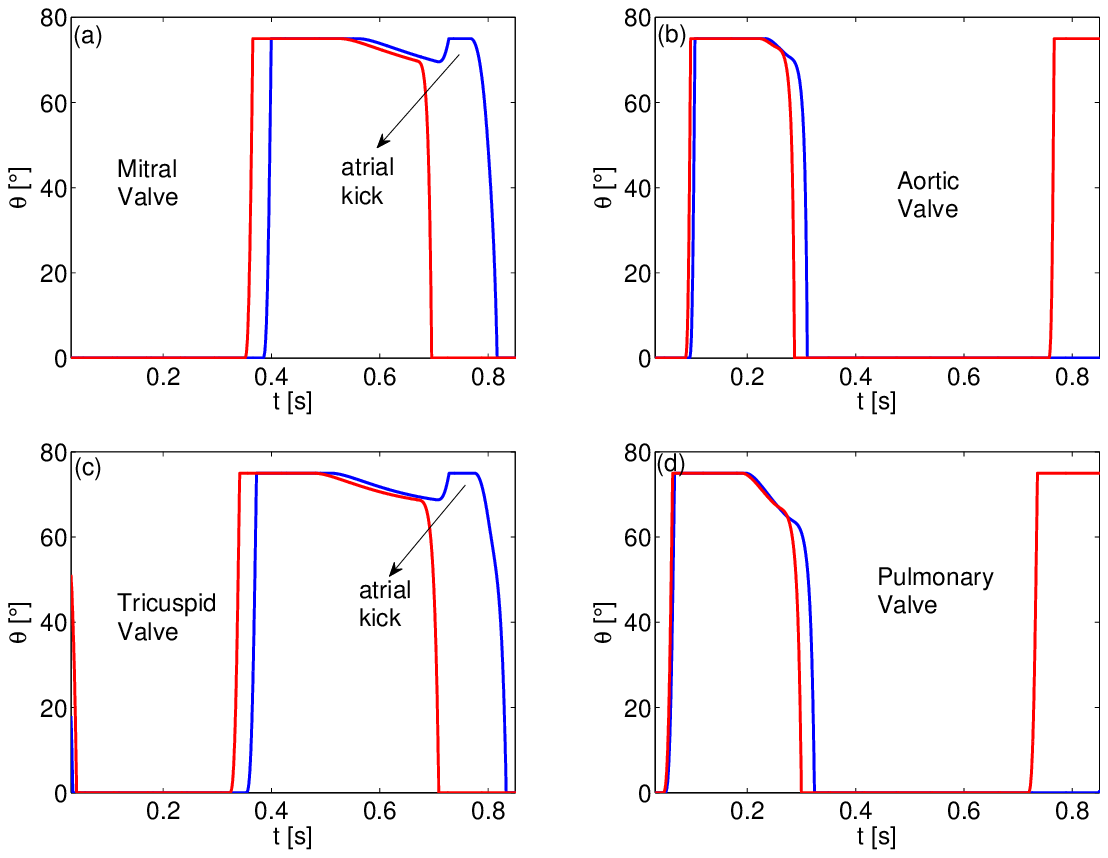}
\renewcommand{\caption}{\footnotesize Figure B1.}
\caption{}
\label{Appendix_B}
\end{figure}


\begin{thebibliography}{12}

\bibitem[Aboelkassem et~al.(2015)]{Aboelkassem} Aboelkassem Y, Savic D, Campbell SG. 2015. Mathematical modeling of aortic valve dynamics during systole. J Theor Biol 365:280-8.

\bibitem[Anselmino et~al.(2015)]{anselmino2015} Anselmino M, Scarsoglio S, Camporeale C, Saglietto A, Gaita F, Ridolfi L. 2015. Rate Control Management of Atrial Fibrillation: May a Mathematical Model Suggest an Ideal Heart Rate? PLoS ONE 10(3): e0119868.

\bibitem[Baccani et~al.(2003)]{baccani} Baccani B, Domenichini F, Pedrizzetti G. 2003. Model and influence of mitral valve opening during the left ventricular filling. J Biomech 36(3):355-61.

\bibitem[Cawley et~al.(2009)]{Cawley} Cawley PJ, Maki JH, Otto CM. 2009. Cardiovascular Magnetic Resonance Imaging for Valvular Heart Disease
Technique and Validation. Circulation 119: 468--478.

\bibitem[Dujardin et~al.(1999)]{Dujardin} Dujardin KS, Enriquez-Sarano M, Schaff HV, Bailey KR, Seward JB, Tajik AJ. 1999. Mortality and morbidity of aortic regurgitation in clinical practice. A long-term follow-up study. Circulation 99(14):1851--7.

\bibitem[Enriquez-Sarano et~al.(1995)]{Enriquez-Sarano1995} Enriquez-Sarano M, Miller FA Jr, Hayes SN, Bailey KR, Tajik AJ, Seward JB. 1995. Effective mitral regurgitant orifice area: clinical use and pitfalls of the proximal isovelocity surface area method. J Am Coll Cardiol 25(3):703--9.

\bibitem[Enriquez-Sarano and Sundt(2010)]{Enriquez-Sarano2010} Enriquez-Sarano M, Sundt TM 3rd. 2010. Is early surgery recommended for mitral regurgitation? Circulation 121(6):804--11.

\bibitem[Fuster et~al.(2006)]{Fuster} Fuster V, Ryden LE, Cannom DS, Crijns HJ, Curtis AB, Ellenbogen KA et al. 2006. ACC/AHA/ESC 2006 Guidelines for the management of patients with atrial fibrillation. Circulation 114: e257--354.

\bibitem[Gertz et~al.(2011)]{Gertz} Gertz ZM, Raina A, Saghy L, Zado ES, Callans DJ, Marchlinski FE, et al. 2011. Evidence of atrial functional mitral regurgitation due to atrial fibrillation: reversal with arrhythmia control. J Am Coll Cardiol 58(14):1474--81.

\bibitem[Grigioni et~al.(2002)]{grigioni} Grigioni F, Avierinos JF, Ling LH, Scott CG, Bailey KR, Tajik AJ, et al. 2002. Atrial fibrillation complicating the course of degenerative mitral regurgitation: determinants and long-term outcome. J Am Coll Cardiol 40(1):84--92.

\bibitem[Kalmanson et~al.(1975)]{Kalmanson} Kalmanson D, Bernier A, Veyrat C, Witchitz S, Savier CH, Chiche P. 1975. Normal pattern and physiological significance of mitral valve flow velocity recorded using transseptal directional Doppler ultrasound catheterization. Br Heart J 37(3):249--56.

\bibitem[Korakianitis and Shi(2006)]{Korakianitis} Korakianitis T, Shi Y.  2006. Numerical simulation of cardiovascular dynamics with healthy and diseased heart valves. J Biomech 39:1964--1982.

\bibitem[{\"O}zdemir et~al.(2001)]{Ozdemir} {\"O}zdemir K, Altunkeser BB, S{\"o}kmen G, Toka\c{c} M, G{\"o}k H. 2001. Usefulness of peak mitral inflow velocity to predict severe mitral regurgitation in patients with normal or impaired left ventricular systolic function. Am Heart J 142(6):1065--71.

\bibitem[Roes et~al.(2009)]{Roes} Roes SD, Hammer S, van der Geest RJ, Marsan NA, Bax JJ, Lamb HJ, et al. 2009. Flow assessment through four heart valves simultaneously using 3-dimensional 3-directional velocity-encoded magnetic resonance imaging with retrospective valve tracking in healthy volunteers and patients with valvular regurgitation. Invest Radiol 44(10):669--75.

\bibitem[Scarsoglio et~al.(2014)]{scarsoglio2014} Scarsoglio S, Guala A, Camporeale C, Ridolfi L. 2014. Impact of atrial fibrillation on the cardiovascular system through a lumped-parameter approach. Med Biol Eng Comput 52(11):905--920.

\bibitem[Sechtem et~al.(2014)]{Sechtem} Sechtem U, Pflugfelder PW, Cassidy MM, White RD, Cheitlin MD, Schiller NB, et al. 1988. Mitral or aortic regurgitation: quantification of regurgitant volumes with cine MR imaging. Radiology 167(2):425--30.

\bibitem[Sharma et~al.(2012)]{Sharma} Sharma S, Lardizabal J, Monterroso M, Bhambi N, Sharma R, Sandhu R, et al. 2012. Clinically unrecognized mitral regurgitation is prevalent in lone atrial fibrillation. World J Cardiol 4(5): 183–-187.

\bibitem[Thavendiranathan et~al.(2012)]{Thavendiranathan} Thavendiranathan P, Phelan D, Collier P, Thomas JD, Flamm SD, Marwick TH. 2012. Quantitative assessment of mitral regurgitation: how best to do it. JACC Cardiovasc Imaging 5(11):1161--75.

\bibitem[Thomas et~al.(1998)]{Thomas} Thomas L, Foster E, Schiller NB. 1998. Peak mitral inflow velocity predicts mitral regurgitation severity. J Am Coll Cardiol 31(1):174--9.

\bibitem[Touche et~al.(1985)]{Touche} Touche T, Prasquier R, Nitenberg A, de Zuttere D, Gourgon R. 1985. Assessment and follow-up of patients with aortic regurgitation by an updated Doppler echocardiographic measurement of the regurgitant fraction in the aortic arch. Circulation 72(4):819--24.

\bibitem[Weinberg et~al.(2010)]{Weinberg} Weinberg EJ, Shahmirzadi D, Mofrad MR. 2010. On the multiscale modeling of heart valve biomechanics in health and disease. Biomech Model Mechanobiol 9(4):373-87.

\bibitem[Yamasaki et~al.(2006)]{Yamasaki} Yamasaki N, Kondo F, Kubo T, Okawa M, Matsumura Y, Kitaoka H, et al. 2006. Severe tricuspid regurgitation in the aged: atrial remodeling associated with long-standing atrial fibrillation. J Cardiol 48(6):315--23.

\bibitem[Zhou et~al.(2002)]{Zhou} Zhou X, Otsuji Y, Yoshifuku S, Yuasa T, Zhang H, Takasaki K, et al. 2002. Impact of atrial fibrillation on tricuspid and mitral annular dilatation and valvular regurgitation. Circ J 66(10):913--6.

\bibitem[Zoghbi et~al.(2003)]{Zoghbi} Zoghbi WA, Enriquez-Sarano M, Foster E, Grayburn PA, Kraft CD, Levine RA, et al. 2003. Recommendations for evaluation of the severity of native valvular regurgitation with two-dimensional and Doppler echocardiography. J Am Soc Echocardiogr 16(7):777--802.

\end{thebibliography}
\end{document}